\documentclass[12pt,a4paper]{article}
\usepackage[english]{babel}
\usepackage{amsmath,amsfonts,amssymb,amsthm}
\newcommand{\cK}{\mathcal{K}}
\newcommand{\cS}{\mathcal{S}}
\newcommand{\cH}{\mathcal{H}}

\newcommand{\cE}{\mathcal{E}}

\begin{document}

\title{A general  proof of Landauer-B\"uttiker formula}
\author{Gheorghe Nenciu\\ 
Dept.  Theor. Phys., Univ. of Bucharest\\ P.O. Box MG 11, RO-077125, 
Bucharest, Romania\\
and \\ 
Institute of Mathematics of the Romanian Academy\\  
PO Box 1-764, 
RO-014700 Bucharest, Romania. \\
E-mail Gheorghe.Nenciu@imar.ro}
\date{}
\maketitle
 
\begin{abstract}

We point out a general argument leading from the formula for currents
through an open noninteracting mesoscopic system given by the theory of non-equilibrium
steady states (NESS) to  the Landauer-B\"uttiker formula.
\end{abstract}
 
Landauer-B\"uttiker type formulas i.e. expressions relating the
(charge, energy etc) currents through mesoscopic systems connected
with electron reservoirs to the corresponding transmission
coefficients have been proved to be a key tool for
analyzing the quantum conductance in nanostructures. Obtained initially
for the stationary case by phenomenological arguments \cite {BILP},
\cite{B}, \cite{IL} they have been widely extended and used. As for the
derivation, one usually assumes that the reservoirs have a lead
geometry and in order to make use of the asymptotic form of the
scattered state the current is evaluated far away from the scatterer
\cite{BS}, \cite{FL}, \cite{MPD}, a
procedure justified (at least in the stationary regime) by charge
conservation. However this approach can become problematic for other
reservoir geometries when the leads are short or even inexistent (see
e.g. \cite{HDQEB}, \cite{BBMH}) or in non-stationary regime.

At a more basic level one starts from a non-equilibrium statistical
mechanics formulation (e.g. linear response theory, NESS theory etc)
and the problem of proving the Landauer-B\"uttiker formula is to show
that the obtained formula for the current can be cast in a form in
which the structure of the mesoscopic system enters only via its
transition matrix of the associated scattering problem as suggested by
the phenomenological derivation. To be more
specific let us consider the NESS procedure (see \cite{AJPP} and
references therein). One starts at $t=0$ from an equilibrium state of
the {\em decoupled} system (i.e. no coupling between the mesoscopic
system and the reservoirs) with reservoirs having different
temperatures and/or chemical potentials. At the one particle level the
system is described by
\begin{equation}\label{decmodel}
H_{0}=H_{\cS}+\sum_{j=1}^{N}H_{j};\cH= \cH_{\cS}\oplus_{j=1}^{N}\cH_{j}
\end{equation}
 where $\cH_{\cS},\;
H_{\cS}$ are the Hilbert space and hamiltonian of the mesoscopic
system and $\cH_{j},\;H_{j}$ are the Hilbert space and the hamiltonian
of the jth reservoir. The coupling, described at the one particle
level by $V$, is switched on suddenly at $t=0$. In the limit
$t\rightarrow \infty$ the system settles down to a non-equilibrium stationary state
\cite{AJPP}. The currents out from the reservoirs are defined as minus
the time
variation of their charge. Since the electrons are considered
independent the second quantization machinery allows to write the
currents in terms of one-particle objects. More precisely
if $\beta_{j},\;\mu_{j}$ are the temperature and the chemical
potential respectively of the jth reservoir in the initial state then
the current out from the kth reservoir in the ``final'' steady state
is \cite{AJPP} (the charge of the electron is $-e$):
\begin{equation}\label{curent}
j_{k}=ieTr_{\cH}(\Omega_{+}\Pi_{0}F_{0}\Pi_{0}\Omega_{+}^{*}[V,\Pi_{k}])
=ieTr_{\cH}(\Pi_{0}F_{0}\Pi_{0}\Omega_{+}^{*}[V,\Pi_{k}]\Omega_{+})
\end{equation}
where $\Pi_{j}$ are the orthogonal projections onto $\cH_{j}$ in
$\cH$, $\Pi_{0}=\sum_{j=1}^{N}\Pi_{j}$,
\begin{equation}\label{moller}
\Omega_{
+}=s-\lim_{t\rightarrow
 - \infty}e^{it(H_{0}+V)}e^{-itH_{0}}\Pi_{0}
\end{equation}
 (we follow the notation  in the physical literature and \cite{RS}) and
\begin{equation}\label{initial}
\Pi_{0} F_{0}\Pi_{0}=\sum_{j=1}^{N}f_{\beta_{j},\mu_{j}}^{FD}(H_{j}).
\end{equation}

An alternative way (and more satisfactory from the physical point of
view) of computing currents in non-equilibrium statistical mechanics is
to start at $t=-\infty$ with reservoirs at {\em the same} temperature and chemical
potential and with an  equilibrium state of the {\em coupled}
system and  then switch on adiabatically the bias in chemical
potential and/or temperature. Unfortunately due to the fact that in
this case the ``perturbation'' is not localized the problem is much
more difficult and it has been worked out only at the linear response
theory level. In this context the Landauer-B\"uttiker formula has been
shown to hold true at the heuristic level by Baranger and Stone
\cite{BS}
and rigorously proved for a tight-binding model for reservoirs by
Cornean, Jensen and Moldoveanu \cite{CJM}.

Coming back to the formula (\ref{curent}) the problem is that the
M{\o}ller operator, $\Omega_{+}$, involves only ``half'' of the
evolution from $-\infty$ to $\infty$ encoded in the scattering
matrix so one has to show that one can rewrite the current only in
terms of scattering matrix and the initial equilibrium state.
 In the related context of adiabatic quantum pumps theory it has been
proved in  \cite{AEGSS} that this is indeed the case  for the lead geometry
of the reservoirs.
Also the fact that  for the exactly solvable  Wigner-Weisskopf
Atom model (\ref{curent}) leads to Landauer-B\"uttiker
formula follows
directly from the results in \cite{AJPP} and the detailed proof
appeared in \cite{JKP}.

The aim of this note is a narrow one:  to outline a general argument
leading from (\ref{curent}) to the Landauer-B\"uttiker formula.
The
argument is entirely elementary and very general: it works for an
arbitrary geometry of the reservoirs (e.g. half spaces, semi infinite
leads with arbitrary section etc) and arbitrary mesoscopic systems of
finite size. Also we allow the reservoirs to be coupled both via the
mesoscopic system and by direct contacts.
 Actually, the only thing which  is needed is a  good stationary
scattering theory for the pair $(H_{0}, H_{0}+V)$. This is consistent
 with the generality of the phenomenological arguments
leading to Landauer-B\"uttiker formula.

In order not to burden the simplicity of the argument in technical and
notational details we shall give it at the formal level and in a
simple context: two reservoirs with the same simple (single channel) absolutely
continuous spectrum, $\sigma_{0} \subset [0, \infty)$,  and mesoscopic systems with a finite number of
states (i.e. $ dim\; \cH_{\cS}=M < \infty$). At the end of the note we
give some straightforward extensions of (\ref{LB}) to more general situations.
 For a mathematical
substantiation one has to make precise the technical conditions on
$H_{0}$ and $V$ and then check that one can apply the results  of the
rigorous stationary scattering theory as developed e.g. in \cite{AJS},
\cite{Y1}, \cite{Y2} (for stationary scattering theory at the formal
level we send the reader to \cite{GW}).

We suppose that the spectral representation of $H_{j}$  $j=1,2$ is given in
terms of generalized eigenfunctions, $|\psi_{j,E}^{0}>$,  living in an
appropriate ``weighted'' Hilbert space,  $\cK_{j}^{*}$ (Gelfand triplets
structure: $\cK_{j} \subset \cH_{j} \subset \cK_{j}^{*}$) :
\begin{equation}
H_{j}|\psi_{j,E}^{0}>=E|\psi_{j,E}^{0}>,\; E\in \sigma(H_{j})=\sigma_{0} \subset [0,\infty).
\end{equation}
For $f\in \cH_{j}$ we denote by $f(E)$ its generalized Fourier
transform:
\begin{equation}\label{gft}
f(E)=<\psi_{j,E}^{0},f>.
\end{equation}

As concerning $V$,  we suppose to have the following structure (in the
decomposition given by (\ref{decmodel})):
\begin{equation}\label{V}
    V=\left(%
\begin{array}{ccc}
  0 & V_{\cS 1} & V_{\cS 2}\\
  V_{\cS 1}^{*} & 0 &  V_{12}\\
   V_{\cS 2}^{*}& V_{12}^{*}& 0
\end{array}%
\right).
\end{equation}
Since $\cH_{\cS}$ has finite dimension, $ V_{\cS 1}$ and $ V_{\cS 2}$
are finite rank operators. We suppose the ``direct contact'', $
V_{12}$,  to be also of finite rank. Accordingly:
\begin{equation}\label{cuplaj1}
 V_{\cS j}=\sum_{l=1}^{m_{j}\leq M}v_{jl}|s_{jl}><f_{jl}|,
\end{equation}
\begin{equation}\label{cuplaj2}
 V_{12}=\sum_{l=1}^{m < \infty}v_{l}|g_{1l}><g_{2l}|
\end{equation}
where $j=1,2$ ,  $\left\{f_{jl}\right\}_{l=1}^{m_{j}}$,
$\left\{g_{jl}\right\}_{l=1}^{m}$,
$\left\{s_{jl}\right\}_{l=1}^{m_{j}}$ are orthonormal systems in
$\cH_{j}$ and  $\cH_{\cS}$ respectively and $v_{jl},\;v_{l}>0$.

Since $V$ is of finite rank, by Kato-Kuroda-Birman theory \cite{RS},
\cite{Y1}, \cite{Y2} the M{\o}ller
operators
\begin{equation}\label{mollerpm}
\Omega_{\pm}=s-\lim_{t\rightarrow \mp \infty}e^{it(H_{0}+V)}e^{-itH_{0}}\Pi_{0}
\end{equation}
exist and are unitary from $\Pi_{0}\cH$ onto the absolutely continuous
subspace,
$\Pi_{ac}\cH$, of $H$.

We impose further conditions on $f_{jl}$, $g_{jl}$ in  order to assure that $\Omega_{\pm}$ provide spectral representations for $H$
restricted to $\Pi_{ac}\cH$ ie for all $E\in \sigma_{ac}(H)=
\cup_{j=1}^{n}\sigma(H_{j})$,
with a possible exception of a discrete set, $\cE$,  $\Omega_{\pm}$
have bounded extensions in the orthogonal sum of $\cK_{j}^{*}$ and
\begin{equation}
|\psi_{j,E}^{\pm}>=\Omega_{\pm}|\psi_{j,E}^{0}>
\end{equation}
are generalized eigenfunctions for $H$:
\begin{equation}
H|\psi_{j,E}^{\pm}>=E|\psi_{j,E}^{\pm}>.
\end{equation}
 A sufficient condition (which at the price of more technicalities can be
weakened) in the case when $H_{j}$ are discrete or continuous
Laplaceans supplemented with boundary conditions is that
$f_{jl}$, $g_{jl}$ are exponentially localized in space. This
condition also
implies that the generalized Fourier coefficients,
$f_{jl}(E)$, $g_{jl}(E)$ (see (\ref{gft}))  of $f_{jl}$, $g_{jl}$ are
smooth functions of $E$, a fact which is needed in order to apply the principal
value formula during the proof below.
The generalized eigenfunctions satisfy the Lippmann-Schwinger
\cite{GW},\cite{AJS}, \cite{Y1}, \cite{Y2}
equation:
\begin{equation}\label{LS}
|\psi_{j,E}^{\pm}>=|\psi_{j,E}^{0}>-(H_{0}-E \mp i0)^{-1}V|\psi_{j,E}^{\pm}>.
\end{equation}

Consider now the scattering operator
\begin{equation}\label{S}
S=\Omega_{-}^{*}\Omega_{+}
\end{equation}
and the corresponding transition operator, $T$, defined by
\begin{equation}
S=1-2i\pi T.
\end{equation}

Since $S$ (and then $T$) commutes with $H_{0}$ it has a  spectral
representation:
\begin{equation}\label{TE}
S=\int_{\sigma_{0}}S(E) dE,\;\; 
T=\int_{\sigma_{0}}T(E) dE
\end{equation}
where $S(E)$ is a unitary two by two matrix (we are considering the
case of  two
reservoirs with simple spectrum). From the unitarity of $S(E)$ it
follows that $T(E)$ satisfies the so called optical theorem:
\begin{equation}\label{toptica}
T(E)-T^{*}(E)=-2\pi i T(E)T^{*}(E).
\end{equation}
The basic result of the stationary scattering theory is the formula
for $T(E)$ in terms of the generalized eigenfunctions of $H_{0}$
\cite{GW}, \cite{AJS}, \cite{Y1}, \cite{Y2}:
\begin{equation}\label{Tpsi} 
T_{jk}(E)=<\psi_{j,E}^{0},V\psi_{k,E}^{+}>=
<\psi_{j,E}^{0},V\Omega_{+}\psi_{k,E}^{0}>.
\end{equation}

To prove the Landauer-B\"uttiker formula in the context described
above amounts to prove that:
\begin{eqnarray}\label{LB}
&j_{1}=ieTr_{\cH}(\Omega_{+}\Pi_{0}F_{0}\Pi_{0}\Omega_{+}^{*}[V,\Pi_{1}])=
\nonumber
\\
&-
2e\pi\int_{\sigma_{0}} dE(f_{\beta_{1},\mu_{1}}^{FD}(E)-
f_{\beta_{2},\mu_{2}}^{FD}(E)
)|T_{12}(E)|^{2}.
\end{eqnarray}
The second equality in (\ref{LB}) is
 the main result of this note.

We compute  $j_{1}$  from (\ref{curent}). Inserting the formula for
$V$ (see (\ref{V}), (\ref{cuplaj1}), (\ref{cuplaj2})) and computing the
trace in appropriate bases one gets:
\begin{eqnarray}\label{curent1}
&j_{1}=-2e\Im (\sum_{l=1}^{l=m_{1}}v_{1l}<f_{1l},
\Omega_{+}\Pi_{0}F_{0}\Pi_{0}\Omega_{+}^{*}s_{1l}>+
\nonumber
\\
&
\sum_{l=1}^{l=m}v_{l}<g_{1l},
\Omega_{+}\Pi_{0}F_{0}\Pi_{0}\Omega_{+}^{*}g_{2l}>).
\end{eqnarray}
 Using the
spectral representation of $\Pi_{0}F_{0}\Pi_{0}$ (see (\ref{initial})) 
in (\ref{curent1}) or, alternatively, evaluating directly the trace in the
r.h.s. of (\ref{curent}) in the generalized basis of $H_{0}$ one gets:
\begin{eqnarray}\label{curent2}
& j_{1}=-
2e \int_{\sigma_{0}}
dE\{f_{\beta_{1},\mu_{1}}^{FD}(E)\Im 
<V\Omega_{+}\psi_{1,E}^{0},\Pi_{1}\Omega_{+}\psi_{1,E}^{0}>+
\nonumber
\\
&
f_{\beta_{2},\mu_{2}}^{FD}(E)\Im <V\Omega_{+}\psi_{2,E}^{0},\Pi_{1}\Omega_{+}\psi_{2,E}^{0}>
\}.
\end{eqnarray}
Let us compute first the coefficient of
$f_{\beta_{1},\mu_{1}}^{FD}(E)$ in (\ref{curent2}).
Using the Lippmann-Schwinger equation (see (\ref{LS})) for
$\Omega_{+}\psi_{1,E}^{0}$ and
 the spectral representation of $H_{0}$ one has:
\begin{eqnarray}\label{fp0}
&<V\Omega_{+}\psi_{1,E}^{0},\Pi_{1}\Omega_{+}\psi_{1,E}^{0}>=
\nonumber
\\
& 
<V\Omega_{+}\psi_{1,E}^{0},\psi_{1,E}^{0}>-
<V\Omega_{+}\psi_{1,E}^{0},\Pi_{1}\frac{1}{H_{0}-E-i0}V\Omega_{+}\psi_{1,E}^{0}>=
\nonumber
\\
& 
\overline{T_{11}(E)}-
\int dE'\frac{| <\psi_{1,E'}^{0},V\Omega_{+}\psi_{1,E}^{0}>|^{2}}{E'-E-i0}.
\end{eqnarray}
Now
the important fact is that we need only the imaginary part of
(\ref{fp0}).
Then using the principal value formula 
\begin{equation}\label{PV}
\frac{1}{x-i0}= i\pi \delta(0)+ PV\frac{1}{x},
\end{equation}
to evaluate the integral in (\ref{fp0})  ( since 
$| <\psi_{1,E'}^{0},V\Omega_{+}\psi_{1,E}^{0}>|^{2}$ depends upon $E'$
only via $f_{1l}(E')$ and  $g_{1l}(E')$ which are smooth by assumption,
this is legitimate)
 one obtains that the coefficient of
$f_{\beta_{1},\mu_{1}}^{FD}(E)$ in (\ref{curent2}) is
$$ \Im (\overline{T_{11}(E)}-i\pi |T_{11}(E)|^{2}).$$  Now  the  use of the optical
theorem (\ref{toptica}) leads to the conclusion that the coefficient of
$f_{\beta_{1},\mu_{1}}^{FD}(E)$ in (\ref{curent2}) is $\pi |T_{12}(E)|^{2}$.
A similar computation for the coefficient  of $f_{\beta_{2},\mu_{2}}^{FD}(E)$ in (\ref{curent2})
 (in this case the term linear in $T(E)$ vanishes) leads to
 $-\pi|T_{12}(E)|^{2}$ 
and the proof of (\ref{LB}) is finished.

We end up with a few remarks:

i. A  similar proof applied to the energy
current (see e.g. \cite{AJPP}) gives:
\begin{eqnarray}\label{LB1}
&\Phi_{1}=-iTr_{\cH}(\Omega_{+}\Pi_{0}F_{0}\Pi_{0}\Omega_{+}^{*}[V,\Pi_{1}H_{0}\Pi_{1}])=
\nonumber
\\
&
2\pi\int_{\sigma_{0}} dE(f_{\beta_{1},\mu_{1}}^{FD}(E)-
f_{\beta_{2},\mu_{2}}^{FD}(E)
)E|T_{12}(E)|^{2}.
\end{eqnarray}

ii. 
The condition that $\sigma (H_{1})=\sigma (H_{2})$ (as  sets) is not
necessary ; in the general case only the energies in the intersection
of $\sigma
(H_{1})$ with $\sigma (H_{2})$ can have nontrivial scattering and then
contribute to the current.

iii.  The straightforward generalization of (\ref{LB}) to the case of
$N$ reservoirs is given by:
\begin{eqnarray}\label{LBN}
&j_{k}=ieTr_{\cH}(\Omega_{+}\Pi_{0}F_{0}\Pi_{0}\Omega_{+}^{*}[V,\Pi_{k}])=
\nonumber
\\
&-
2e\pi\int_{\sigma_{0}} dE\sum_{j=1}^{N}(f_{\beta_{k},\mu_{k}}^{FD}(E)-
f_{\beta_{j},\mu_{j}}^{FD}(E)
)|T_{
kj}(E)|^{2}.
\end{eqnarray}
Notice that for $N>2$  and  systems without time reversal symmetry one can have $|T_{
kj}(E)|^{2}\neq |T_{
jk}(E)|^{2}$.  Still $\sum_{k=1}^{N}j_{k}=0$, as required from charge
conservation, due to the fact that the unitarity of $S$ implies that
$T(E)T^{*}(E)=T^{*}(E)T(E)$.

iv. Finally, if at some energy $E$ the spectra of $H_{j}$ do not have
multiplicity one then $T_{jk}(E)$  become operators
$T_{jk}(E):\cH_{j}(E) \rightarrow \cH_{k}(E)$ and $|T_{
jk}(E)|^{2}$ in (\ref{LBN})  are to be   replaced by  $Tr_{\cH_{j}(E)}T_{jk}(E)T_{jk}^{*}(E)$.

{\bf Acknowledgments}

This research has been supported  by the CEEX Grant
05-D11-45/2005.  The final
version of this note has been written  during a visit at Aalborg
University. Both the financial support and the hospitality of the
Department of Mathematical Sciences, Aalborg University, are
gratefully acknowledged.

\vspace{5mm}

{\em Note added} : After the first version of this note has been archived,
Claude-Alain Pillet informed me that a detailed proof of the
Landauer-Buttiker formula for the Wigner-Weisskopf Atom model appeared
in \cite{JKP}. I would like also to thank Claude-Alain Pillet 
 for many
stimulating discussions about NESS theory in general and about the
derivation of the Landauer-B\"uttiker
formula in particular.

\end{document}